\documentclass[reprint,aps,pra,amsmath,amssymb]{revtex4-1}

\usepackage[utf8]{inputenc}
\usepackage[T1]{fontenc}
\usepackage{graphicx}
\usepackage[hidelinks]{hyperref}
\usepackage{mathtools}
\usepackage{braket}
\DeclarePairedDelimiter{\abs}{\lvert}{\rvert}

\begin{document}

\title{Collision model for non-Markovian quantum trajectories}

\author{S. J. Whalen}
\email{simon.whalen@gmail.com}
\affiliation{The Dodd-Walls Centre for Photonic and Quantum Technologies, Department of Physics, University of Auckland, Private Bag 92019, Auckland, New Zealand}

\begin{abstract}
We present an algorithm to simulate genuine, measurement-conditioned quantum trajectories for a class of non-Markovian systems, using a collision model for the environment. We derive two versions of the algorithm, the first corresponding to photodetection and the second to homodyne detection with a finite local oscillator amplitude. We use the algorithm to simulate trajectories for a system with delayed coherent feedback, as well as a system with a continuous memory.
\end{abstract}

\maketitle

\section{Introduction}\label{sec:introduction}

\vspace*{-1.75mm}A quantum trajectory is a sequence of states of an open quantum system, conditioned on a sequence of measurements of the system's output. Markovian quantum trajectory theory describes ``unravelings'' of a Lindblad master equation: each unraveling is a decomposition of the system density matrix corresponding to a specific measurement set-up~\cite{carmichael1993}. This approach leads to a family of efficient Monte Carlo algorithms that can be used to solve the dynamics of Markovian open quantum systems. It has been shown that pure-state quantum trajectories for non-Markovian systems do not exist in general~\cite{wiseman_pure-state_2008}, but this does not preclude the existence of mixed-state trajectories for such systems. While various Monte Carlo methods for analyzing non-Markovian systems have been proposed~\cite{strunz_linear_1996,diosi_non-markovian_1997,diosi_non-markovian_1998,strunz_open_1999,breuer_genuine_2004,piilo_non-markovian_2008,megier_parametrization_2018}, there remains---in contrast to the Markovian case---no generally applicable computational tool that produces genuine measurement-conditioned trajectories for systems with environmental memories. One way around this is to model part of the environment along with the system, as any non-Markovian system can be rendered Markovian by simply enlarging the system. This is the approach we follow in this paper.

Collision models, in which the dynamics of open quantum systems are modeled through repeated unitary interactions between the system and constituent subsystems of the environment~\cite{campbell_system-environment_2018}, are emerging as a general tool for the understanding of non-Markovian systems~\cite{kretschmer_collision_2016,ciccarello_collision_2017}. In recent years, models of this kind have been applied to the problem of coherent feedback---a signature example of an environmental memory---by \citet{grimsmo_time-delayed_2015} and separately by \citet{pichler_photonic_2016}. Collision models are particularly convenient as a computational tool: by simulating a portion of the environment along with the system, these models are able to treat memory effects while avoiding some of the challenges associated with analytical approaches. A connection between models of this kind and quantum trajectory theory was pointed out by \citet{brun_simple_2002}, who used a collision model (although this terminology was not in use at the time) as a simple illustration of the essential physics of Markovian, pure-state quantum trajectories. Quantum trajectory theory has also been used in conjunction with collision models by Cresser to analyze the micromaser~\cite{cresser_micromaser_2006,cresser_time-reversed_2019}, and by Dąbrowska and coauthors to model systems where non-Markovianity arises due to correlations in the input field~\cite{dabrowska_quantum_2017,dabrowska_quantum_2019,dabrowska_quantum_2019-1}. In this paper we show how to use a collision model to simulate quantum trajectories of a class of non-Markovian open systems with environmental memories. The algorithm we derive provides a fairly efficient and general method of simulating the dynamics of these systems.

In Sec.~\ref{sec:model} we introduce a collision model for an open quantum system, which in general can be non-Markovian. In Sec.~\ref{sec:trajectory-algorithm} we describe a quantum trajectory algorithm within this model, focusing on trajectories for two different detection schemes: photodetection (Sec.~\ref{sec:photodetection}) and homodyne detection (Sec.~\ref{sec:homodyne-traj}). In Sec.~\ref{sec:delay-coher-feedb} we apply this algorithm to a specific example of a non-Markovian quantum system, namely a system with delayed coherent feedback, and also illustrate how the approach can be generalized to systems with continuous memories. We conclude in Sec.~\ref{sec:conclusion}\@. Natural units in which $\hbar = c = 1$ are used throughout.

\vspace*{-5mm}\section{Model}\label{sec:model}

\vspace*{-1.75mm}Our starting point is a simple, prototypical open quantum system that interacts with a one-dimensional, freely propagating electromagnetic field. This set-up has been studied elsewhere~\cite{grimsmo_time-delayed_2015,pichler_photonic_2016}, but we recapitulate the basic model here. The system is allowed to couple to the environment at multiple spatial locations, which in general can make the system's dynamics non-Markovian, as we discuss in detail in Sec.~\ref{sec:delay-coher-feedb}. Sampling the environment at discrete intervals leads to a collision model that approximately describes the evolution of the combined system.

We model the environment on an interval $[-L, 0]$. We denote the frequency domain annihilation operators for an environment mode by $b_k$, while the corresponding time-domain operators are denoted by $B_n$. We sample the time domain at equal intervals $\Delta t = L/N$, so that these operators are related by the discrete Fourier transform:
\begin{align}
  \label{eq:1}
  b_k &= \frac{1}{\sqrt{N}} \sum_{n=0}^{N-1} B_n e^{-i \omega_k n \Delta t} \,, \notag\\ \Longleftrightarrow \quad B_n &= \frac{1}{\sqrt{N}} \sum_{k=0}^{N-1} b_k e^{i \omega_k n \Delta t} \,,
\end{align}
where $\omega_k = 2 \pi k / L$. Because we are using natural units, the time domain operator $B_n$ can also be thought of as representing the environment field between positions $-n \Delta t$ and $-(n+1) \Delta t$.

We now introduce a system with creation and annihilation operators $a^\dagger$ and $a$. The Hamiltonian for the combined system can be split up as
\begin{equation}
  \label{eq:2}
  H = H_S + H_E + H_I \,.
\end{equation}
Here $H_S$ describes the internal dynamics of the system, while
\begin{equation}
  \label{eq:3}
  H_E = \sum_{k=0}^{N-1} \omega_k b_k^\dagger b_k
\end{equation}
generates the free evolution of the environment. The interaction between system and environment is described, in the rotating-wave approximation, by
\begin{equation}
  \label{eq:4}
  H_I = \sum_{k=0}^{N-1} \left( \kappa_k a^\dagger b_k + \kappa_k^* b_k^\dagger a \right) \,,
\end{equation}
where
\begin{equation}
  \label{eq:5}
  \kappa_k = \frac{1}{\sqrt{L}} \sum_{n=0}^{N-1} \gamma_n e^{i \omega_k n \Delta t} \,,
\end{equation}
with $\gamma_n$ being a position-dependent coupling strength. Substituting $\kappa_k$ from Eq.~\eqref{eq:5} into Eq.~\eqref{eq:4} and reexpressing in terms of the time domain operators $B_n$ gives
\begin{equation}
  \label{eq:6}
  H_I = \frac{1}{\sqrt{\Delta t}} \sum_{n=0}^{N-1} \left( \gamma_n a^\dagger B_n + \gamma_n^* B_n^\dagger a \right) \,.
\end{equation}
We use this form of the interaction Hamiltonian in what follows.

The evolution of the system is described by the unitary operator $U(t) = e^{-i H t}$. For small $\Delta t$ we can make the approximation
\begin{equation}
  \label{eq:7}
  U(\Delta t) \approx \Delta U_{E} \Delta U_{SI} \,,
\end{equation}
where $\Delta U_E = e^{-i H_E \Delta t}$ and $\Delta U_{SI} = e^{-i (H_S + H_I) \Delta t}$. Equation~\eqref{eq:7} leads to an algorithm for approximating the evolution of the system and environment. First, the state is evolved with $H_S + H_I$ through a time $\Delta t$, which is equivalent to applying $\Delta U_{SI}$. The free evolution of the environment is then accounted for by applying $\Delta U_{E}$. Using $e^{i H_E t} b_k e^{-i H_E t} = b_k e^{-i \omega_k t}$, we can see that
\begin{equation}
  \label{eq:8}
  \Delta U_E^\dagger B_n \Delta U_E = \frac{1}{\sqrt{N}} \sum_{k=0}^{N-1} b_k e^{i \omega_k (n - 1) \Delta t} = B_{n-1} \,,
\end{equation}
showing that excitations in the environment propagate in the direction of decreasing $n$.

\begin{figure}
  \centering
  \includegraphics[page=1]{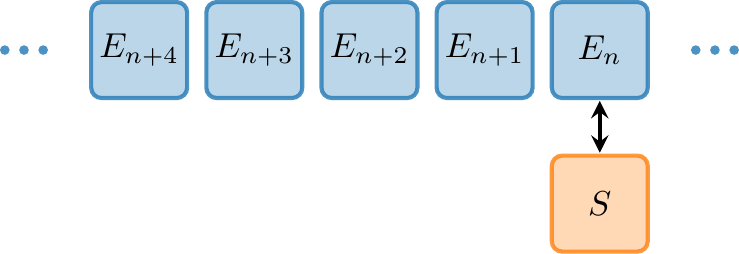}

  \vspace*{0.65cm}

  \includegraphics[page=2]{figures}
  \caption{Two time steps of the unitary collision model described in Sec.~\ref{sec:model}. In the first step (top) the system interacts with the $n$\textsuperscript{th} environment subsystem for a time $\Delta t$. Between the first and second steps, the unitary $\Delta U_E$ is applied, which as indicated by Eq.~\eqref{eq:10} can be thought of as translating the environment to the right. As such, in the second step (bottom) the system interacts with the $(n+1)$\textsuperscript{st} subsystem. The open-system dynamics depicted in this diagram are Markovian, because the system $S$ couples to the environment at a single location.}\label{fig:collision-schematic-basic}
\end{figure}

To make the effect of $\Delta U_E$ more explicit, we fix a basis for the environment by defining the number states
\begin{equation}
  \label{eq:9}
  \ket{k_{M-1}, \ldots, k_0} = \ket{k_{M-1}} \cdots \ket{k_0} \,,
\end{equation}
where $0 < M \leq N$, and where $\ket{k_n}$ is a number state of a single environment oscillator. We can express any state of the environment, which we recall comprises $N$ oscillators, as a linear combination of the states $\ket{k_{N-1}, \ldots, k_0}$. Equation~\eqref{eq:8} then leads to
\begin{equation}
  \label{eq:10}
  \Delta U_E \ket{k_{N-1}, \ldots, k_1, k_0} = \ket{k_0, k_{N - 1}, \ldots, k_1} \,.
\end{equation}
We thus arrive at a collision model describing our system. In this picture, the system interacts stroboscopically with an environment comprising a chain of harmonic oscillators whose annihilation operators are $B_n$. In the first stage of the evolution, the system interacts with these environment oscillators according to the Hamiltonian $H_S + H_I$ for a time interval $\Delta t$. In the second stage, the unitary operator $\Delta U_E$ is applied; as Eq.~\eqref{eq:10} shows, this shifts any excitations in the $n$\textsuperscript{th} oscillator into the $(n-1)$\textsuperscript{st} for $n > 0$, while excitations in the zeroth oscillator end up in the $(N-1)$\textsuperscript{st}. These stages are iterated to produce $\Delta t$-periodic samples of the system state. Two time steps of this evolution are illustrated, for a Markovian system, in Fig.~\ref{fig:collision-schematic-basic}.

\section{Trajectory algorithm}\label{sec:trajectory-algorithm}

\begin{figure*}
  \centering
  \includegraphics[page=3]{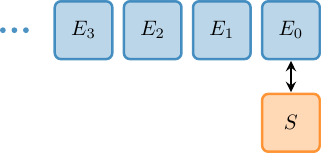}
  \hspace*{0.6cm}
  \includegraphics[page=4]{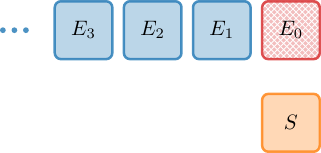}
  \hspace*{0.6cm}
  \includegraphics[page=5]{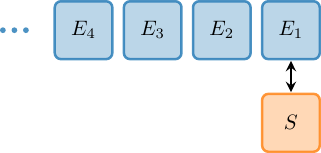}
  \caption{Schematic illustration of the trajectory algorithm described in Sec.~\ref{sec:trajectory-algorithm}, for a Markovian open system. From left to right: first, the system interacts with the environment, represented by a collision model, for a time $\Delta t$; second, a simulated measurement is performed on the zeroth environment oscillator, disentangling it from the rest of the combined system; third, the environment oscillators are shifted along by one step, and interaction with the system resumes. }\label{fig:collision-schematic-trajectory}
\end{figure*}

Due to the periodic boundary condition implicit in Eq.~\eqref{eq:1}, the model presented in the previous section will eventually begin to display unphysical behavior, as excitations in the zeroth oscillator ``loop around'' to the $(N-1)$\textsuperscript{st}. We can avoid this by simulating a measurement of the zeroth environment oscillator to disentangle this oscillator from the system and the rest of the environment. The state of the zeroth oscillator can then be ignored in future time steps, thereby creating an absorbing boundary condition at the zeroth oscillator and allowing the simulation to continue indefinitely.

There are two ways in which this measurement can be represented. We can take the partial trace over the zeroth subsystem, representing a probability-weighted sum over the possible measurement outcomes, which would yield a density-matrix representation of the state of the system and environment. Alternatively, we can perform a Monte Carlo simulation of the application of the Born rule. Here we focus on the latter approach, which leads to a quantum trajectory unraveling of the system--environment combined state. Trajectories for the system alone can then be obtained by tracing out the environment, which will in general be a mixed-state trajectory. These trajectories for the system alone provide an unraveling of the system density matrix, in the same sense as the conventional algorithm does for Markovian systems, and are contextual: the specific decomposition of the density matrix obtained from the algorithm depends on the measurement set-up being simulated.

Our algorithm, illustrated in Fig.~\ref{fig:collision-schematic-trajectory},  involves three steps, which are iterated to obtain a discrete-time quantum trajectory for the combined system:
\begin{enumerate}
\item\label{item:step1} Apply $\Delta U_{SI}$ by evolving the state of the combined system with the Hamiltonian $(H_S + H_I)$ through a time $\Delta t$.
\item\label{item:step2} Make a simulated measurement of some observable of the zeroth environment oscillator.
\item\label{item:step3} Apply a truncated form of $\Delta U_E$ to the postmeasurement state.
\end{enumerate}
Step~\ref{item:step1} is easily performed by using any standard differential equation solver. We used \textsc{ZVODE} from \textsc{ODEPACK}~\cite{odepack}, interfaced through SciPy~\cite{scipy}. We consider steps~\ref{item:step2} and~\ref{item:step3} in more detail below.

Step~\ref{item:step2} of our algorithm is to make a simulated measurement of some observable of the zeroth environment oscillator using the Born rule. The specific observable to be measured determines what kind of quantum trajectory we obtain. Two examples are considered in Secs.~\ref{sec:photodetection} and~\ref{sec:homodyne-traj}, but for the time being we keep things generic. Consider an observable of the zeroth oscillator with discrete eigenstates $\{\ket{q_j}\}$, corresponding to eigenvalues $\{q_j\}$. The probability of each measurement result is easily calculated as $P(q_j) = \abs{\braket{q_j|\psi}}^2$, where $\ket{\psi}$ is the state of the system and environment. A standard weighted pseudorandom choice is used to determine the measurement outcome at each step, and the zeroth oscillator is then projected into some fiducial state $\ket{\varphi}$: $\ket{\psi} \to \ket{\varphi}\braket{q_j|\psi}$. The choice of fiducial state is influenced by our chosen basis, which we discuss below.

Step~\ref{item:step3} of our algorithm is to apply a truncated form of $\Delta U_E$, denoted $\widetilde{\Delta U}_E$, which is defined analogously to Eq.~\eqref{eq:10}. The states $\ket{k_{N-1}, \ldots, k_1} \ket{\varphi}$ form a basis for the system after it has undergone the simulated measurement process described above, with the zeroth oscillator projected into the fiducial state $\ket{\varphi}$. We therefore define $\widetilde{\Delta U}_E$ by its action on these basis states:
\begin{equation}
  \label{eq:11}
  \widetilde{\Delta U}_E \ket{k_{N-1}, \ldots, k_1} \ket{\varphi} = \ket{0, k_{N-1}, \ldots, k_1} \,.
\end{equation}
The effect of applying $\widetilde{\Delta U}_E$ is therefore to shift excitations in the $n$\textsuperscript{th} oscillator into the $(n-1)$\textsuperscript{st}, while ``resetting'' the $(N-1)$\textsuperscript{st} to its vacuum state. We can easily construct a suitable $\widetilde{\Delta U}_E$:
\begin{equation}
  \label{eq:12}
  \widetilde{\Delta U}_E = \sum_{\mathclap{k_{N-1}, \ldots, k_1}} \ket{0, k_{N-1}, \ldots, k_1} \bra{k_{N-1}, \ldots, k_1} \bra{\varphi} \,.
\end{equation}
Alternatively, if $\ket{\varphi}$ is chosen to be the vacuum state, the unitary $\Delta U_E$ in fact already satisfies Eq.~\eqref{eq:11}.

Finally, we need to consider the truncation of the environment Hilbert space. Our environment comprises $N$ harmonic oscillators, and we have chosen to represent these in a number-state basis. Our chosen truncation is to constrain the values of $k_n$ that appear in Eq.~\eqref{eq:9}. In particular, we choose $k_n \leq 1$, so that each environment oscillator effectively becomes a qubit, and require the total number of excitations in the environment not to exceed some number: $\sum_n k_n \leq K_{\text{max}}$. As such, the basis states of the environment are all the combinations of $0, 1, \ldots, K_{\text{max}}$ excitations. The simulations presented in this paper were performed with $K_{\text{max}} = 2$ except where otherwise specified.

We have yet to specify exactly what is measured in step~\ref{item:step2} of our algorithm. In the following two subsections, we consider two different measurement schemes, and show in each case that our collision model reduces to the corresponding Markovian quantum trajectory algorithm in the appropriate limit.

\subsection{Photodetection trajectories}\label{sec:photodetection}

Perhaps the simplest measurement scheme is to simulate detection of the excitation number in the zeroth oscillator. We show here that for a Markovian system this corresponds to the well-known quantum trajectory theory of direct photodetection.

For a Markovian system it is sufficient to consider a single environment oscillator, which means that the interaction Hamiltonian~\eqref{eq:6} simplifies to
\begin{equation}
  \label{eq:13}
  H_I = \frac{1}{\sqrt{\Delta t}} \left( \gamma a^\dagger B + \gamma^* B^\dagger a \right) \,.
\end{equation}
As we have chosen to truncate the environment Hilbert space such that each oscillator contains at most one excitation, $B$ is the lowering operator for a qubit. Coherent evolution through $\Delta t$ followed by a measurement performed on the environment qubit represents a single time step of our collision model. Suppose the initial state of the combined system is $\ket{\psi(0)} = \ket{\psi_S} \ket{0}$. For small $\Delta t$, we can make the approximation
\begin{equation}
  \label{eq:14}
  \ket{\psi(\Delta t)} \approx \left( 1 - \frac{\Delta t }{2} \abs{\gamma}^2 a^\dagger a \right) \ket{\psi_S} \ket{0} - i \sqrt{\Delta t} \gamma^* a \ket{\psi_S} \ket{1} \,,
\end{equation}
where we have neglected any internal dynamics of the system for simplicity. The excitation of the environment qubit $B^\dagger B$ is then measured: the result $1$ is identified with a ``click'' in a photodetector, while the result $0$ is identified with no detection. From Eq.~\eqref{eq:14}, we can immediately see that the probability of such a click is $\abs{\gamma}^2 \braket{\psi_S | a^\dagger a | \psi_S} \Delta t$, and that the normalized system state after such a detection will be $a \ket{\psi_S} / \sqrt{\braket{\psi_S | a^\dagger a | \psi_S}}$, up to a phase factor. The state conditioned on no detection is $\left( 1 - \frac{\Delta t }{2} \abs{\gamma}^2 a^\dagger a \right) \ket{\psi_S}$, ensuring conservation of probability to first order in $\Delta t$. These are exactly the results of the conventional quantum trajectory theory of direct photodetection~\cite{carmichael1993,dalibard_wave-function_1992}.

\subsection{Homodyne trajectories}\label{sec:homodyne-traj}

We now wish to extend our trajectory treatment to homodyne detection. In this case, the system and environment are augmented with an ancillary system, the local oscillator. This is a harmonic oscillator prepared, at each time step, in a coherent state $\ket{\alpha \sqrt{\Delta t}}$. The Hamiltonian describing the coherent evolution of the combined system, in the Markovian case, is again Eq.~\eqref{eq:13}. Note that the local oscillator does not evolve under this Hamiltonian because it is already prepared in the desired state. Coherent evolution through $\Delta t$, followed by a joint measurement on the zeroth environment qubit and local oscillator, represents a single step of the stroboscopic evolution of a collision model.

Balanced homodyne detection consists in simultaneously measuring two environment observables, $N_+$ and $N_-$, where
\begin{equation}
  \label{eq:15}
  N_\pm = \frac{1}{2} {\left(B \pm C\right)}^\dagger \left(B \pm C\right) \,,
\end{equation}
and where $B$ and $C$ are annihilation operators for the zeroth environment qubit and the local oscillator respectively. The eigenstates of these operators are $\ket{n\pm} = (\ket{0, n} \pm \ket{1, n - 1}) / \sqrt{2}$ with $n \geq 1$, as well as the vacuum. We find
\begin{equation}
  \label{eq:16}
  N_+ \ket{n\pm} = \lambda_{n\pm} \ket{n\pm} \quad \text{and} \quad N_- \ket{n\pm} = \lambda_{n\mp} \ket{n\pm} \,,
\end{equation}
where $\lambda_\pm = (n \pm \sqrt{n})/2$. If $\ket{\alpha}$ is a coherent state and $\ket{\varphi}$ is some generic state of the qubit, we have the inner product $\braket{n\pm|\varphi, \alpha} = e^{-\abs{\alpha}^2/2} \alpha^{n-1} ( \alpha \braket{0|\varphi} \pm \sqrt{n} \braket{1|\varphi} ) / \sqrt{2 n!}$.

We now aim to show that this measurement scheme is equivalent to homodyne detection, in the relevant parameter regime. The argument proceeds much as it did for photodetection. The initial state is now $\ket{\psi(0)} = \ket{\psi_S} \ket{0, \alpha \sqrt{\Delta t}}$, meaning that Eq.~\eqref{eq:14} becomes
\begin{multline}
  \label{eq:17}
  \ket{\psi(\Delta t)} \approx \left( 1 - \frac{\Delta t }{2} \abs{\gamma}^2 a^\dagger a \right) \ket{\psi_S} \ket{0, \alpha \sqrt{\Delta t}} \\ - i \sqrt{\Delta t} \gamma^* a \ket{\psi_S} \ket{1, \alpha \sqrt{\Delta t}} \,.
\end{multline}
Homodyne detection corresponds to a measurement in the $\ket{n\pm}$ basis, and provided $\Delta t$ is small enough the probability of observing states with $n > 1$ will be negligible. We note that $N_\pm \ket{1\pm} = \ket{1\pm}$, while $N_\pm \ket{1\mp} = 0$. As such, we have three eigenstates to consider: $\ket{0, 0}$, the vacuum state of the environment, corresponding to no detection; $\ket{1+}$, corresponding to a click in the $N_+$ detector; and $\ket{1-}$, corresponding to a click in the $N_-$ detector. To first order in $\Delta t$, the corresponding inner products are
\begin{align}
  \braket{0, 0|\psi(\Delta t)} &= \left[ 1 - \frac{\Delta t}{2} \left( \abs{\alpha}^2 + \abs{\gamma}^2 a^\dagger a \right) \right] \ket{\psi_S} \,, \label{eq:18} \\
  \braket{1{\pm}|\psi(\Delta t)} &= J_{\pm} \ket{\psi_S} \sqrt{\Delta t} \,, \label{eq:19}
\end{align}
where we have defined
\begin{equation}
  \label{eq:20}
  J_{\pm} = \frac{\alpha \mp i \gamma^* a}{\sqrt{2}} \,.
\end{equation}
These operators are, up to a phase convention, the jump operators that appear in the conventional trajectory algorithm for homodyne detection with finite local oscillator amplitude~\cite{carmichael2008p455}. Finally, we arrive at the probabilities
\begin{align}
  P_0 &= 1 - \left( \abs{\alpha}^2 + \abs{\gamma}^2 a^\dagger a \right) \Delta t \,, \label{eq:21} \\
  P_\pm &= \braket{\psi_S | J_\pm^\dagger J_\pm | \psi_S} \Delta t \,. \label{eq:22}
\end{align}
We note in particular that Eq.~\eqref{eq:22} is exactly what we would expect from the conventional trajectory algorithm. From this, we can conclude that the measurement scheme described above does indeed correspond to a trajectory simulation of homodyne detection, in the limit of small $\Delta t$.

A further simplification is possible: in a balanced homodyne detection scheme, the results of the $N_-$ measurement are subtracted from the results of the $N_+$ measurement. As such we can, equivalently, measure the operator $Q = N_+ - N_- = C^\dagger B + B^\dagger C$. We can see that $Q \ket{n\pm} = \pm \sqrt{n} \ket{n\pm}$, along with, of course, $Q \ket{0, 0} = 0 \ket{0, 0}$. Provided we only see eigenvalues with $n \leq 1$ we can interpret this in the same way as the conventional homodyne algorithm: we have three possible measurement outcomes ($\pm1$ or $0$) representing a click in each of the two detectors, or neither. This way of expressing the problem is computationally convenient because we only need to simulate measurement in a single output channel, where previously we had two.

\begin{figure}
  \centering
  \includegraphics[page=6]{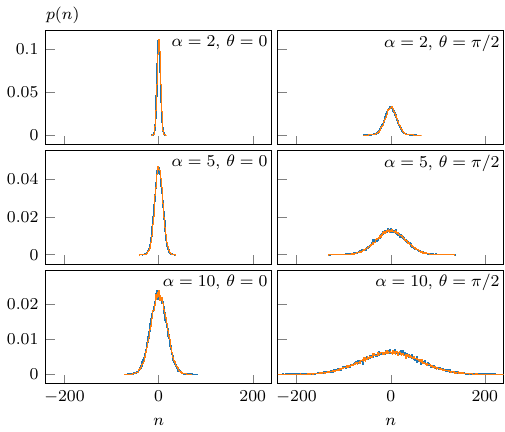}
  \caption{Comparison of ``counting'' distributions for the homodyne detection of squeezed light. The system is a harmonic oscillator with Hamiltonian $H_S = i \zeta ({a^\dagger}^2 - a^2)$, and the local oscillator field has amplitude $\alpha e^{i\theta}$. Orange lines (front) show results from the conventional trajectory algorithm with jump operators given by Eq.~\eqref{eq:20}; with this phase convention, $\theta=0$ corresponds to detection of the squeezed $Y$ quadrature and $\theta=\pi/2$ corresponds to the unsqueezed $X$ quadrature. Blue lines (back) correspond to an analogous collision model simulation, with $\gamma \Delta t=0.01$ and up to $249$ excitations in the local oscillator. The horizontal axis is the sum of measurements in an interval of length $10/\gamma$, after a ``burn-in'' time of $4/\gamma$ that reduces the influence of transient dynamics. The squeezing parameter is given by $\zeta=0.1\gamma$, and the system Hilbert space is truncated at nine excitations. Each histogram was prepared from 50,000 trajectories.}\label{fig:squeezing}  
\end{figure}

For the equivalence between our collision model and a conventional trajectory simulation of homodyne detection to hold, we need to choose a small enough time step that there is at most one click in each interval. For this to be the case, we must decrease $\Delta t$ quadratically as we increase $\alpha$, such that $\alpha \sqrt{\Delta t}$ remains below some threshold. This is not computationally feasible, so for larger values of $\alpha$ we inevitably end up performing simulations with large enough $\Delta t$ that we observe the eigenvalues corresponding to $n \geq 2$, which will in general not be integers. Nonetheless, numerical comparisons to the conventional jump algorithm, such as those illustrated in Fig.~\ref{fig:squeezing}, indicate that the correspondence remains approximately valid for a range of $\alpha$.

\section{Delayed coherent feedback}\label{sec:delay-coher-feedb}

\begin{figure}
  \centering
  \includegraphics[page=7]{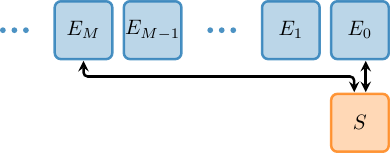}
  \caption{Collision model of an open quantum system with delayed coherent feedback. The system (yellow, below) couples to two different environment subsystems (blue, above), creating a feedback loop with delay $\tau = M \Delta t$.}\label{fig:collision-schematic-delay}
\end{figure}

In Sec.~\ref{sec:trajectory-algorithm} we showed that our algorithm reproduces the results of conventional quantum trajectory theory for Markovian systems, when applied to both photodetection and homodyne detection. We now turn our attention to non-Markovian systems. The parameters $\gamma_n$ that appear in Eq.~\eqref{eq:6} describe position-dependent coupling of the system to the environment. Coupling at a single point, as considered in Sec.~\ref{sec:trajectory-algorithm}, leads to a Markovian open quantum system. If, on the other hand, the system couples to the environment at more than one location, the system can create excitations in the environment that interact again with the system at a later time. Put another way, the environment ``remembers'' the state of the system and feeds this information back coherently after some time delay. Here we consider perhaps the simplest example of such an environmental memory, wherein the system couples to the environment at exactly two spatial locations, creating a coherent environmental feedback loop with a discrete propagation delay. This kind of feedback has previously been studied in work on ``atomic'' emission in front of a mirror~\cite{dorner_laser-driven_2002,carmele_single_2013,tufarelli_dynamics_2013,tufarelli_non-markovianity_2014} and in solid-state systems with significant propagation delays~\cite{guo_giant_2017}. Figure~\ref{fig:collision-schematic-delay} illustrates a collision model for this set-up.

\begin{figure}
  \centering
  \includegraphics[page=8]{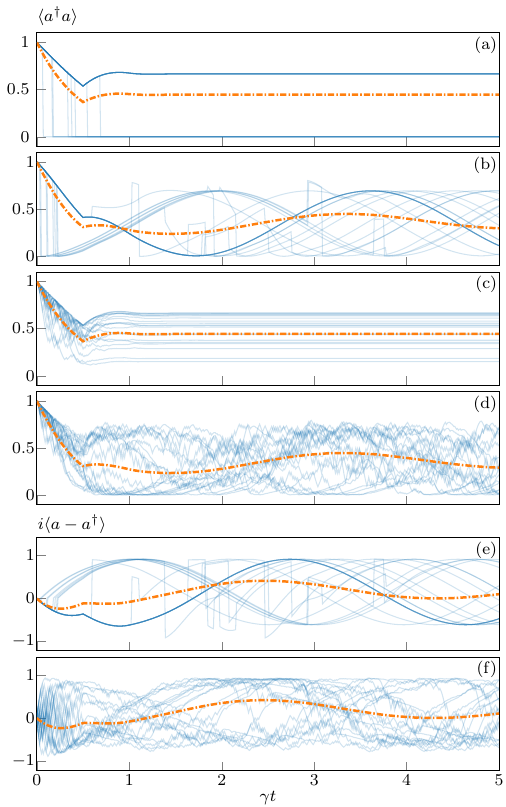}
  \caption{Trajectory simulations of a qubit with delayed coherent feedback. The upper four plots show the occupation number: (a), (b) for photodetection, and (c), (d) for homodyne detection with $\alpha^2 = 100\gamma$. Panels (a) and (c) correspond to emission from an initially excited system without driving, while panels (b) and (d) correspond to a driven qubit with $\Omega/\gamma = 1$. Twenty sample trajectories are plotted as solid blue lines for each configuration. The opacity of the blue curves provides an indication of how common a given trajectory is. Note in particular the trajectories without jumps that appear in darker blue in panels (a) and (b). The corresponding ensemble averages (25,000 trajectories) are shown as dotted orange lines. Trajectories of the Pauli $Y$ operator are shown in panels (e) and (f), corresponding exactly to the driven cases (b) and (d); the onset of the feedback is visible in the ensemble average at $t = \tau$. The other parameters are $\phi = \pi$, $\gamma\tau = 0.5$, $\gamma\Delta t = 0.01$, and local oscillator dimension 250.}\label{fig:sample-traj}
\end{figure}

To make things more concrete, consider the interaction Hamiltonian~\eqref{eq:6}, with $\gamma_0 = \gamma e^{i\phi}$, $\gamma_M = \gamma$, and $\gamma_n = 0$ for all other $n$. This coupling describes an environmental feedback loop of length $\tau = M \Delta t$, with a phase advance of $\phi$ in the loop. We choose as an example system a driven qubit, with free Hamiltonian $H_S = \Omega (a^\dagger + a)$ in a frame rotating at the system resonant frequency, and the system initially in its excited state. Sample trajectories, along with ensemble averages, calculated by using our collision model for this system with both photodetection and homodyne detection, are shown in Fig.~\ref{fig:sample-traj}. Without observing the in-loop field, we cannot tell whether a detected photon was emitted directly from the system or via the loop. This ambiguity means that trajectories for this system are, in general, mixed-state trajectories. Note, however, that this is not a property of coherent feedback as such: a detection delay results in mixed-state trajectories even for a system with Markovian dynamics.

\begin{figure}
  \centering
  \includegraphics[page=9]{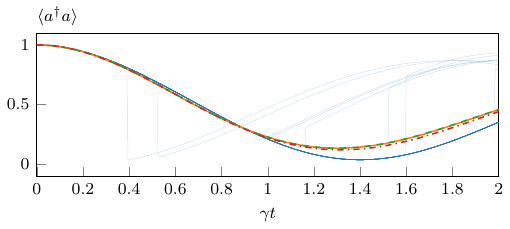}
  \caption{Comparison of simulations with the exponential coupling~\eqref{eq:23}, with $L=5$. Fifty sample photodetection trajectories are plotted as thin blue lines, with the ensemble average (10,000 trajectories) in orange. The broken lines show equivalent simulations using the Markovian master equation for the Jaynes--Cummings model: for the dashed green curve, the cavity Hilbert space was truncated at the single-excitation level, while up to 20 excitations were allowed for the dotted red curve. The trajectory simulations used $K_{\text{max}}=1$, so the ensemble average agrees well with the single-excitation master equation. The multiple-excitation master equation begins to diverge at $\gamma t \approx 1$, as the environment excitation level increases. The other parameters are $\Omega/\gamma=1$, $\lambda/\gamma=1$, and $\gamma\Delta t=0.005$. The size of the environment has been truncated at $L=5\gamma^{-1}$.}\label{fig:jc-traj}
\end{figure}

We thus see that our algorithm is particularly well suited for simulating non-Markovian systems. The main computational advantage is that the algorithm is constant-space, requiring only enough computer memory to store the state of the system along with enough of the environment to model the feedback loop in question. This is in contrast to the approach introduced by \citet{grimsmo_time-delayed_2015} and explored further in Ref.~\cite{whalen_open_2017}, which requires an additional ``copy'' of the system Hilbert space after each delay cycle and thus consumes an amount of memory that grows exponentially with the number of delay intervals to be simulated. The approach adopted by \citet{pichler_photonic_2016}, on the other hand, also has constant space complexity. Our algorithm has the further advantage that it allows us to simulate not just discrete feedback loops but other kinds of quantum memory as well. A continuous environmental memory kernel, where the evolution of the system depends most generally on its state at all previous times, may be approximated as a series of discrete feedback loops. This can be accomplished in our algorithm by choosing appropriate $\gamma_n$. Take as an example the exponential coupling
\begin{equation}
  \label{eq:23}
  \gamma_n = \sqrt{\gamma} \lambda \Delta t e^{-\lambda n \Delta t} \,,
\end{equation}
with $\lambda > 0$. Substituting this coupling into Eq.~\eqref{eq:5} and taking the continuum limit ($L \to \infty$) results in a Lorentzian spectral density $J(\omega) = {(2\pi)}^{-1} \gamma \lambda^2 / (\lambda^2 + \omega^2)$. This spectral density maps to a system coupled to a damped harmonic oscillator initially in the vacuum state~\cite{garraway_nonperturbative_1997,lorenzo_composite_2017}; if the system is a qubit then this corresponds to the Jaynes--Cummings model~\cite{jaynes_comparison_1963}. As a proof of principle Fig.~\ref{fig:jc-traj} shows sample trajectories obtained with an exponential coupling and the environment truncated at a finite ``length'', and compares the ensemble average to results obtained using the Markovian master equation.

\section{Conclusion}\label{sec:conclusion}

In this paper we introduced an algorithm to simulate quantum trajectories for non-Markovian systems, by using a collision model to represent the environment and its interaction with the system. The algorithm produces trajectory unravelings of the system density matrix. As is the case in the well-known Markovian theory of quantum trajectories, these unravelings are contextual in the sense that they depend on the measurement set-up. We provided two versions of the algorithm, corresponding to photodetection and homodyne detection respectively, and illustrated the algorithm's application to two signature examples of non-Markovian dynamics: a coherent feedback loop with a discrete delay, and the Lorentzian spectral density that arises in the Jaynes--Cummings model.

Our simulation method has computational advantages over some alternative approaches. In particular, while the algorithm requires us to simulate a portion of the environment---the ``memory''---in addition to the system, the Hilbert-space dimension of this environmental memory remains constant irrespective of the simulation time, meaning that the algorithm has constant space complexity. The algorithm acts on a pure state of the system and environment, which consumes less computer memory than a density-matrix representation---an advantage our approach shares with stochastic approaches derived from Markovian quantum trajectory theory.

Because our approach is derived directly from a fundamental model of a system interacting with an assemblage of harmonic oscillators, it has in principle very wide applicability when used as a brute force method, given enough computing power. An efficient implementation, however, requires a sufficiently compact representation of the part of the environment containing the memory. The systems studied in this paper have finite-length memories that are only ever populated with one or two photons. Systems with very long memories, or systems whose timescales require a very small step size $\Delta t$, may pose computational difficulties for this method because the number of environment subsystems would be large. The same is true of environments with more than one spatial dimension. Performance issues could also arise with systems that scatter many photons into the environment, necessitating a large $K_{\text{max}}$. In these cases, alternative methods of ``compressing'' the environment state may be required. One possibility is to adapt a matrix product state approach, as used for example by~\citet{pichler_photonic_2016}, for use in quantum trajectory simulations.

Quantum trajectory theory is of interest beyond its use as a numerical tool. Genuine quantum trajectories provide an accurate description of the evolution of a system, conditioned on a sequence of observations of the system's output. The contextuality of a specific trajectory unraveling, depending as it does on the measurement scheme chosen by the observer, has been described as ``subjective reality'' in the context of quantum measurement theory~\cite{wiseman_quantum_1996}. The algorithm presented in this paper generates genuine, measurement-conditioned quantum trajectories for a fairly large class of non-Markovian open quantum systems, namely those where the environment can be represented as a collision model with a position-dependent coupling. This connection to the theory of quantum measurement opens up the possibility of analyzing non-Markovian systems from a new perspective.

\begin{acknowledgments}
The simulations presented in this paper were performed with the aid of the Python library QuTiP~\cite{qutip2}. The author would like to acknowledge Howard Carmichael for his guidance and many fruitful suggestions, and, through him, the support of the Dodd-Walls Centre for Photonics and Quantum Technologies. The author also gratefully acknowledges valuable discussions with Jim Cresser.
\end{acknowledgments}

\bibliography{references}

\end{document}